\documentclass[fleqn,12pt]{article}
\usepackage{amssymb}
\usepackage{amsfonts,amsmath}
\usepackage{latexsym}
\begin{document}

\baselineskip=22pt
\vspace*{1.9cm}

\begin{center}
{\LARGE    The Laplacian on  homogeneous spaces}
\end{center}
\vspace*{0.5cm}

\begin{center}
{\large Liangzhong Hu } \\

\vspace*{0.5cm}
{\small{\it Department of Mathematics, Federal University of Paran\'{a}}} \\
{\small{\it     C. P. 019081, Curitiba, PR, 81531-990,  Brazil}} \\
\end{center}
\vspace*{0.5cm}

\begin{abstract}
The  solution of the eigenvalue problem of the Laplacian on a general 
homogeneous space $G/H$ is given. Here $G$ is a compact, semi-simple Lie group, $H$ is a closed
subgroup of $G$, and the rank of $H$ is equal to the rank of $G$.   
It is shown that the multiplicity of the lowest eigenvalue of the Laplacian on $G/H$ is just the degeneracy of the lowest
Landau level for a particle moving on $G/H$ in the presence of the background gauge field.
Moreover, the eigenspace of the lowest eigenvalue of the Laplacian on $G/H$ is, up to a sign, equal to the $G$-equivariant index of the Kostant's Dirac operator on $G/H$.  
\end{abstract}

\newpage

\section*{I. INTRODUCTION}

Homogeneous spaces play an important role in superstring theories, M-theory, (see, for example 
\cite{ramond}), and the higher dimensional quantum Hall effect \cite{Zhang}.

The higher dimensional quantum Hall effect involves a generalization of the Landau problem for a particle moving on homogeneous spaces in the presence of the background gauge fields. 
Its further development can be found in \cite{Fab,Kar,Ber,Dolan,Hu,Bel,meng,Jel,Nair}.
A common ingredient of these analysis is the calculation of the 
degeneracy of the ground state for the Landau problem.

Since the Landau Hamiltonian on a homogeneous space is, up to a constant, equal to the Laplacian on the same homogeneous 
space, it is meaningful to study the solution of the eigenvalue problem of the Laplacian, especially the multiplicity of the lowest eigenvalue of the Laplacian.
   
In this paper, we give the solution of the eigenvalue problem of the Laplacian on a general homogeneous space $G/H$, where the rank of $H$ equals the rank of ${\bf g}$.

It is shown that the multiplicity of the lowest eigenvalue of the Laplacian on $G/H$ is just the degeneracy of the lowest
Landau level on $G/H$ , although the eigenvalues of the Laplacian are different from the Landau levels.

In \cite{Dolan} Dolan showed how the degeneracy of the lowest Landau level is related to the Atiyah-Singer index theorem for the Dirac operator on a homogeneous space. 

In fact, the Atiyah-Singer index theorem on a homogeneous space reduces to the Weyl dimension formula. More precisely, 
we find that the eigenspace of the lowest eigenvalue of the Laplacian on $G/H$ is, up to a sign, equal to the 
$G$-equivariant index of the Kostant's Dirac operator on $G/H$ \cite{Bot,Kos,land}.  

The layout of the paper is as follows. In section II, we review the gauge field and the Laplacian on $G/H$. In section 
III, we first give the eigenvalues of the Laplacian on $G/H$. By using a generalization of the Weyl Character formula 
discovered by Gross, Kostant, Ramond and Sternberg \cite{GKRS}, the explicit expression of the multiplicity of the lowest 
eigenvalue is obtained.

\section*{II. THE GAUGE FIELD AND THE LAPLACIAN ON $G/H$}

In this section, we briefly review the gauge field and the Laplacian on $G/H$. 
 An introduction to the differential geometry on homogeneous spaces can be found  in \cite{Kob1,Kob2}.
The gauge fields on homogeneous spaces were discussed in \cite{Sto,Lan,Lev,meng,Nie}.  

Let $G$ be a compact, semi-simple Lie group with Lie algebra ${\bf g}$. ${\bf g}$ is identified with all left invariant vector fields on $G$. As a vector space ${\bf g}$ is isomorphic to the tangent space $T_{e}(G)$ at the identity $e$ of $G$.
The Cartan-Killing form on ${\bf g}$ gives rise to a canonical metric $( , )$ on $G$.

Let $H$ be a closed  subgroup of $G$ with Lie algebra ${\boldsymbol{\eta}}$. 

We suppose that $G/H$ is reductive, i.e. ${\bf g}$ has an orthognal decomposition
$
{\bf g} = {\boldsymbol{\eta}} \oplus {\bf m}
$
with $[{\boldsymbol{\eta}}, {\bf m}] \subset {\bf m}$ and $[{\bf m}, {\bf m}] \subset {\bf g}$.

The metric  $(, )$ of $G$ induces a natural reductive metric  $( , )_{\bf m}$ on $G/H$.

The principal $H$-bundle $P(G/H, H)$ has a canonical connection one form $A$ on $G$ with
\begin{equation}
A(X) = Pr_{\boldsymbol{\eta}}(X) \equiv X_{\boldsymbol{\eta}}\in {\boldsymbol{\eta}}, 
\end{equation}
for any $X\in{\bf g}$. Here $Pr_{\boldsymbol{\eta}}$ is the orthgonal projection onto ${\boldsymbol{\eta}}$.

The gauge field  on $G$ is identified as the connection one form $A$.
One has
\begin{eqnarray}
Ker(A) &=& {\bf m}, \\
Im(A) &=& {\boldsymbol{\eta}}.
\end{eqnarray}
For any $X\in{\bf m}$, The covariant derivative reads
\begin{equation}
{\nabla}_{X} = X - A(X) = X.
\end{equation}
 
Notice that when $G$ is a subgroup of the group of all invertible $n\times n$ matrices, one can write the connection one form $A$ as
\[
A = Pr_{\boldsymbol{\eta}}({g}^{-1}{\rm d}{g}),
\]
where $g\in G$.

Let $X_{1}, X_{2}, \cdots, X_{n}$ be  an orthonormal basis of ${\bf m}$.  From equation (4), the Laplacian on $G/H$ takes the form
\begin{equation}
\Delta = {\sum}^{n}_{i=1}{\nabla}^{2}_{i}\equiv {\sum}^{n}_{i=1}{\nabla}^{2}_{X_{i}} = {X}_{i}^{2} = C_{2}(G) - C_{2}(H).
\end{equation}
Here $C_{2}(G)$ and $C_{2}(H)$ are the quadratic Casimir elements of $G$ and $H$, respectively.

Let $U$ be an irreducible representation of $H$.
Let $G{\times}_{H} U$ be the associated vector bundle of the principal bundle $P(G/H, H)$.
The space of square integrable sections of $G{\times}_{H} U$ forms a Hilbert 
space, and it decomposes into the direct sum of the eigenspaces of the Laplacian on $G/H$. These eigenspaces are all irreducible representations of $G$, and this induces the following expression for the Laplacian on $G/H$ which was discussed in \cite{Sto,Lan,Lev,Zhang,Kar,Dolan,Hu,Bel} and appears explicitly in Meng's paper \cite{meng}.

{\bf Proposition 1 (Meng)}. {\it The Laplacian on} $G/H$ {\it is} 
\begin{equation}
\Delta = C_{2}(G, \cdot ) - C_{2}(H, U).
\end{equation}
{\it Here} $C_{2}(G, \cdot)$ {\it is the  quadratic Casimir element of} $G$ {\it calculated in an irreducible 
representation of} $G$. $C_{2}(H, U)$ {\it is the quadratic Casimir element of} $H$ {\it calculated in a given irreducible representation} $U$.

\section*{III. THE SOLUTION OF EIGENVALUE PROBLEM OF $\Delta$ ON $G/H$}

In this section, we impose the following conditions and notation on Lie algebras ${\bf g}$ of $G$ and ${\boldsymbol{\eta}}$
of $H$.
Let $\bf{g}$ be a semi-simple Lie algebra  and let ${\boldsymbol{\eta}}$ be a reductive subalgebra of the same rank 
as ${\bf g}$. This means that we can choose a common Cartan subalgebra 
\[
{\bf h}\subset {\boldsymbol{\eta}} \subset {\bf g}.
\]
Let ${\Phi}_{\bf g}$ be the set of roots of ${\bf g}$.
The roots ${\Phi}_{\boldsymbol{\eta}}$ of ${\boldsymbol{\eta}}$  form a subset of the roots of $\bf{g}$, i.e.,
\[
{\Phi}_{\boldsymbol{\eta}} \subset {\Phi}_{\bf g}.
\]
Choosing a positive root system  ${\Phi}^{+}_{\bf g}$ for $\bf{g}$ also determines a positive root system  
${\Phi}^{+}_{\boldsymbol{\eta}}$ for ${\boldsymbol{\eta}}$, where
\[
 {\Phi}^{+}_{\boldsymbol{\eta}} \subset {\Phi}^{+}_{\bf g}.
\]
Let ${\rho}_{\bf{g}} = {\frac{1}{2}}{\sum}_{ {\alpha}\in {\Phi}^{+}_{\bf g}}\alpha$ and  
${\rho}_{\boldsymbol{\eta}} = {\frac{1}{2}}{\sum}_{ {\alpha}\in {\Phi}^{+}_{\boldsymbol{\eta}}}\alpha$ denote half the sum of the positive roots of $\bf{g}$ and ${\boldsymbol{\eta}}$  respectively.

\subsection*{A. Eigenvalues of $\Delta$ on $G/H$}

Let $U_{\mu}$ be a fixed irreducible representation of ${\boldsymbol{\eta}}$ with dominant highest weight $\mu$. 
The Hilbert space of square integrable sections of $ G{\times}_{H} U_{\mu}$ decomposes into the direct sum of the eigenspaces of $\Delta$, which are irreducible representations $V_{\lambda}$ 
of $\bf g$ with dominant highest weights $\lambda$'s. 
By using the Harish-Chandra isomorphism, there is a natural injective homomorphism $i : Z_{\bf g}\rightarrow 
Z_{\boldsymbol{\eta}}$ where $Z_{\bf g}$ and $Z_{\boldsymbol{\eta}}$ are  centers of the enveloping algebras of 
${\bf g}$ and ${\boldsymbol{\eta}}$ respectively.
This means that $C_{2}(G, V_{\lambda})$ and $C_{2}(H, U_{\mu})$ are mutually commuting.
Finding the spectrum of $C_{2}(G, V_{\lambda})$ is a standard content in textbooks on group representations, 
for example \cite{Bar}. 
The construction of $V_{\lambda}$ by the highest weight cyclic eigenvector  is also a standard content in a textbook.
So we omit these unecessary repetitions.
The spectrum of $C_{2}(G, V_{\lambda})$ is 
\begin{equation}
C_{2}({\lambda}) = (\lambda + {\rho}_{\bf g}, \lambda + {\rho}_{\bf g}) - ({\rho}_{\bf g}, {\rho}_{\bf g}).
\end{equation}
The spectrum of $C_{2}(H, U_{\mu})$ is
\begin{equation}
C_{2}({\mu}) = (\mu + {\rho}_{\boldsymbol{\eta}}, \mu + {\rho}_{\boldsymbol{\eta}}) - ({\rho}_{\boldsymbol{\eta}}, {\rho}_{\boldsymbol{\eta}}).
\end{equation}
Thus we have the following result:

{\bf Proposition 2}. {\it The eigenvalue of} $\Delta$ {\it labelled by dominant highest weight} $\lambda$ reads
\begin{equation}
 E_{\lambda} = (\lambda + {\rho}_{\bf g}, \lambda + {\rho}_{\bf g}) - (\mu + {\rho}_{\boldsymbol{\eta}}, \mu + 
{\rho}_{\boldsymbol{\eta}}) - 
({\rho}_{\bf g}, {\rho}_{\bf g}) +({\rho}_{\boldsymbol{\eta}}, {\rho}_{\boldsymbol{\eta}}).
\end{equation}
{\it The multiplicity of the eigenvalue $E_{\lambda}$ is given by the  Weyl dimension formula}: 
\begin{equation}
dim V_{\lambda} = \frac{{\prod}_{{\alpha}\in {\Phi}^{+}_{\bf g}}(\lambda +\rho_{\bf g} , \alpha)}{{\prod}_{{\alpha}\in 
{\Phi}^{+}_{\bf g}}(\rho_{\bf g} , \alpha)}.
\end{equation}

In subsection  {\bf C} we will find that
\begin{equation}
(\lambda + {\rho}_{\bf g}, \lambda + {\rho}_{\bf g}) \ge (\mu + {\rho}_{\boldsymbol{\eta}}, \mu + 
{\rho}_{\boldsymbol{\eta}}). 
\end{equation}

\subsection*{B. A generalization of the Weyl character formula}

In order to determine the multiplicity of the lowest eigenvalue of $\Delta$,
we review a generalization of the Weyl character formula due to Gross, Kostant, Ramond and Sternberg 
(GKRS) \cite{GKRS}.

Let $W_{\bf{g}}$ be the Weyl group  of $\bf{g}$ which acts simply transitively on the Weyl chambers of $\bf{g}$, each of which is 
contained inside a Weyl chamber for ${\boldsymbol{\eta}}$.  Let
\begin{equation}
C \subset W_{\bf{g}}
\end{equation}
denote the set of elements which map the positive Weyl chamber for $\bf{g}$ into the positive Weyl chamber for ${\boldsymbol{\eta}}$.

For any dominant weight $\lambda$ of $\bf{g}$, the weight $\lambda + {\rho}_{\bf{g}}$ lies in the interior of the
positive Weyl chamber for $\bf{g}$.
Hence for each $c\in C$, the element $c(\lambda + {\rho}_{\bf{g}})$ lies in the interior of the positive Weyl chamber
for ${\boldsymbol{\eta}}$, and 
\begin{equation}
c\bullet \lambda : = c(\lambda + {\rho}_{\bf{g}}) - {\rho}_{\boldsymbol{\eta}}
\end{equation}
is a dominant weight for ${\boldsymbol{\eta}}$, and each of these is distinct.

As in section II, let $\bf{m}$ denote the orthogonal complement of ${\boldsymbol{\eta}}$ in $\bf{g}$. 
We thus get a homomorphism of ${\boldsymbol{\eta}}$ into the orthogonal algebra $o({\bf m})$, which is an even dimensional, and hence has two half-spin representations $S^{+}$ and $S^{-}$ considered as ${\boldsymbol{\eta}}$ modules.

In \cite{GKRS}, Gross, Kostant, Ramond and Sternberg prove the following theorem.

{\bf Theorem 3 (GKRS)}.
{\it Let} $V_{\lambda}$ {\it be the irreducible representation of} $\bf{g}$ {\it with highest weight} $\lambda$. {\it The following identity holds in the representation ring} $R({\boldsymbol{\eta}})$:
\begin{equation}
V_{\lambda}\otimes S^{+} - V_{\lambda}\otimes S^{-} = {\sum}_{c\in C} (-1)^{c} U_{c\bullet \lambda}.
\end{equation}
{\it where} $U_{c\bullet\lambda}$ {\it denotes the irreducible representation of} ${\boldsymbol{\eta}}$ {\it with highest weight} $c\bullet\lambda$. 

In the special case that ${\boldsymbol{\eta}}$ equals the Cartan subalgebra ${\bf h}$,  the identity (14) is just the Weyl character formula.

\subsection*{C. The multiplicity of the lowest eigenvalue of $\Delta$}

Given an irreducible representation $U_{\mu}$  of ${\boldsymbol{\eta}}$ with highest weight $\mu$,  from Theorem 3, $\mu$
corresponds to a unique highest weight $\lambda$ of a irreducible representation $V_{\lambda}$ of $\bf g$. 
More precisely, there exists $c\in C$ such that  
$\mu = c\bullet\lambda = c(\lambda +{\rho}_{\bf g})-{\rho}_{\boldsymbol{\eta}}$.
This means that $\lambda = w(\mu +{\rho}_{\boldsymbol{\eta}})-{\rho}_{\bf g}$ for $w = c^{-1}\in W_{\bf g}$.
If $\lambda $ is dominant for ${\bf g}$, the eigenspace of the lowest eigenvalue is 
$V_{w(\mu +{\rho}_{\boldsymbol{\eta}})-{\rho}_{\bf g}}$ with highest weight
$w(\mu +{\rho}_{\boldsymbol{\eta}})-{\rho}_{\bf g}$.
It follows that 
$(\lambda + {\rho}_{\bf g}, \lambda + {\rho}_{\bf g}) = (w(\mu + {\rho}_{\boldsymbol{\eta}}), w(\mu + {\rho}_{\boldsymbol{\eta}})) = (\mu + {\rho}_{\boldsymbol{\eta}}, \mu + {\rho}_{\boldsymbol{\eta}}). $
 By Proposition 2, the lowest eigenvalue is $E_{w(\mu +{\rho}_{\boldsymbol{\eta}})-{\rho}_{\bf g}} = 
({\rho}_{\boldsymbol{\eta}}, {\rho}_{\boldsymbol{\eta}})-({\rho}_{\bf g}, {\rho}_{\bf g}) $.
By using the Weyl dimension formula,
the multiplicity of the lowest eigenvalue can be obtained.
Thus we have the following result:

{\bf Theorem 4}. {\it Given an irreducible representation} $U_{\mu}$ {\it of} 
${\boldsymbol{\eta}}$ {\it with highest weight} $\mu$. {\it If there exists
an element} $w\in W_{\bf g}$  {\it in the Weyl group of} ${\bf g}$ {\it such that the weight} $w(\mu +{\rho}_{\boldsymbol{\eta}})-{\rho}_{\bf g}$ {\it is dominant for} ${\bf g}$.
{\it Then the lowest eigenvalue of} $\Delta$ {\it is} 
\begin{equation}
E_{w(\mu +{\rho}_{\boldsymbol{\eta}})-{\rho}_{\bf g}} 
= ({\rho}_{\boldsymbol{\eta}}, {\rho}_{\boldsymbol{\eta}})-({\rho}_{\bf g}, {\rho}_{\bf g}),
\end{equation}
{\it and the  multiplicity of the lowest eigenvalue of } $\Delta$ {\it is}
\begin{equation}
dim V_{w(\mu +{\rho}_{\boldsymbol{\eta}})-{\rho}_{\bf g}} = \frac{{\prod}_{{\alpha}\in {\Phi}^{+}_{\bf g}}
(w(\mu +{\rho}_{\boldsymbol{\eta}}), \alpha)}{{\prod}_{{\alpha}\in {\Phi}^{+}_{\bf g}}(\rho_{\bf g} , \alpha)}.
\end{equation}
{\bf Remark.}
1. By Theorem 4, we have the inequality (11).

2. If $\lambda = w(\mu +{\rho}_{\boldsymbol{\eta}})-{\rho}_{\bf g}$ is not dominant for ${\bf g}$, the lowest eigenvalue of $\Delta$ does not exist. 
   Thus we can always choose $\mu$ such that $\lambda$ is dominant.

3. $V_{w(\mu +{\rho}_{\boldsymbol{\eta}})-{\rho}_{\bf g}}$ is, up to a sign, equal to the $G$-equivariant index of the 
   Kostant's Dirac operator on $G/H$ \cite{land}.  

{\bf Example}. 
We take $ {\boldsymbol{\eta}} = so(2n)$, ${\bf g} = so(2n+1)$ and make the usual choice of Cartan subalgebra and positive roots,
${\epsilon}_{i}\pm {\epsilon}_{j}$ $(1\le i < j \le n)$ for 
${\boldsymbol{\eta}}$ and these together with ${\epsilon}_{i}$, $(1\le i \le n)$ for ${\bf g}$, the interior of positive Weyl chamber for ${\bf g}$
 consists of all
\[
x = (x_{1}, \cdots , x_{n}) = x_{1}{\epsilon}_{1}+\cdots 
+ x_{n}{\epsilon}_{n}
\]
 with
\[
x_{1} > x_{2}>\cdots >x_{n} > 0,
\]
whereas the interior of positive Weyl chamber of 
${\boldsymbol{\eta}}$
consists of all $x$ satisfying
\[
x_{1} > x_{2}>\cdots >x_{n-1} > \vert x_{n}\vert .
\]
Thus $C$ consists of the identity $e$ and the reflection $s$, which
changes the sign of the last coordinate. Here
\[
{\rho}_{\bf g} = (\frac{2n-1}{2}, \frac{2n-3}{2}, \cdots , \frac{1}{2})
\]
and
\[
{\rho}_{\boldsymbol{\eta}} = (n-1 , n-2 , \cdots , 1 , 0).
\]
One has
\[
{\rho}_{\bf g} - {\rho}_{\boldsymbol{\eta}} = (\frac{1}{2}, \cdots , \frac{1}{2}).
\]
Thus given an irreducible representation $U_{\mu}$ of ${\boldsymbol{\eta}}$ with dominant highest weight $\mu = 
({\mu}_{1},\cdots , {\mu}_{n})$ or $({\mu}_{1},\cdots , -{\mu}_{n})$, the eigenspace of the lowest eigenvalue 
of $\Delta$ on the $2n$ dimensional sphere
$S^{2n} = {\rm SO}(2n+1)/{\rm SO}(2n)$ is 
$V_{\lambda}$ with highest weight
$\lambda = ({\mu}_{1} - \frac{1}{2},\cdots , {\mu}_{n}-\frac{1}{2})$.
The multiplicity of the lowest eigenvalue of $\Delta$ is
\[
dim V_{({\mu}_{1} - \frac{1}{2},\cdots , {\mu}_{n}-\frac{1}{2})}
 = {\prod}^{n}_{i = 1} \frac{2{\mu}_{i} + 2n -2i}{(2n+1- 2i )!}
{\prod}_{1\le i < j \le n} ({\mu}_{i} - {\mu}_{j} +j -i)({\mu}_{i} +{\mu}_{j} + 2n -i-j).
\]
Now we consider two special cases.

1. When $\mu = (\frac{1}{2},\cdots , \frac{1}{2})$ or $(\frac{1}{2},\cdots , \frac{1}{2}, -\frac{1}{2})$, one has
$\lambda= (0,\dots ,0)$. In this special case $S^{+} = U_(\frac{1}{2},\cdots , \frac{1}{2})$ and 
$S^{-} = U_(\frac{1}{2},\cdots , \frac{1}{2}, -\frac{1}{2})$ are the actual half-spin representations of 
$so(2n)$.
The multiplicity of the lowest eigenvalue of $\Delta$ is
\[
dim V_{(0,\cdots , 0)}=
{\prod}_{1\le i < j \le n} ( j -i)(2n+1 -i-j).
\]

2. Given $\mu = (\frac{I}{2},\cdots \frac{I}{2})$ or $(\frac{I}{2},\cdots \frac{I}{2}, -\frac{I}{2})$ with $I$ a natural
number, we have  
$\lambda = (\frac{I - 1}{2},\cdots , \frac{I - 1}{2})$.
The multiplicity of the lowest eigenvalue of $\Delta$ is
\[
dim V_{(\frac{I - 1}{2},\cdots , \frac{I - 1}{2})}
 = {\prod}^{n}_{i = 1} \frac{I + 2n -2i}{(2n+1- 2i )!}
{\prod}_{1\le i < j \le n} (j -i)(I + 2n -i-j).
\]
When $n = 1$, $\mu = \frac{I}{2}$ or $-\frac{I}{2}$, we have $\lambda = \frac{I - 1}{2}$.
The multiplicity of the lowest eigenvalue of $\Delta$  on the $2$-sphere is $dim V_{\frac{I - 1}{2}} = I$.

\section*{IV. DISCUSSION}

We have obtained the solution of the eigenvalue problem of the Laplacian on a general homogeneous space $G/H$, where the rank of $H$ equals the rank of ${\bf g}$.

The Landau Hamiltonian of a particle moving on $G/H$ can be expressed as:
\begin{equation}
\hat{H} = \frac{{\hbar}^{2}}{2M}{\Delta} = \frac{{\hbar}^{2}}{2M} [C_{2}(G, \cdot ) - C_{2}(H, U_{\mu})],
\end{equation}
where $M$ is the inertia mass of the particle.
We argue that the multiplicity of the lowest eigenvalue of $\Delta$ on $G/H$ is exactly the degeneracy of the lowest Landau level on $G/H$.
Thus Theorem 4 provides a general formula to calculate the degeneracy of the lowest Landau level on  general homogeneous spaces.

It should be mentioned that $\lambda = w(\mu +{\rho}_{\boldsymbol{\eta}})-{\rho}_{\bf g}$ is the highest weight of the eigenspace 
of the ground Landau level. In the literature, this highest weight was often taken to be $\lambda = \mu$. However this 
does not affect the physical consequences, if the thermodynamic limit is taken.  

Notice that the eigenvalues of  $\Delta$ are different from the Landau levels due to the different evaluation of the 
quadratic Casimir element $C_{2}(H, U_{\mu})$. In the Landau problem \cite{Zhang,Fab,Kar,Ber,Dolan,Hu,Bel,meng,Jel,Nair}, 
the evaluation of $C_{2}(H, U_{\mu})$ depends on the structure of each $H$. So we can not get an explicit expression
for the Landau levels on general homogeneous spaces.

Moreover, we find that the eigenspace of the lowest eigenvalue of $\Delta$ on $G/H$ is, up to a sign, equal to the $G$-equivariant index of the Kostant's Dirac operator on $G/H$ \cite{Bot,Kos,land}.  
So Theorem 4 provides a convenient method to calculate the index of homogeneous differential operators on $G/H$.

\section*{ ACKNOWLEDGMENTS}

I am grateful to E. Hoefel, R. Raya and A. Sant'Anna for discussions.

\end{document}